\documentclass[preprint, 12pt, a4paper, review]{elsarticle}

\usepackage{amssymb}
\usepackage{hyperref}
\usepackage{tabularx}
\usepackage[normalem]{ulem}
\usepackage{hyphenat}
\usepackage{algorithm}
\usepackage{algorithmic}
\usepackage{listings}
\usepackage{spverbatim}
\usepackage{subcaption}
\usepackage{amsmath}

\newcommand{\MYCOMMENT}[1]{(#1)}

\journal{SoftwareX}

\begin{document}
\renewcommand{\labelenumii}{\arabic{enumi}.\arabic{enumii}}

\begin{frontmatter}


\title{Mamute: high-performance computing for geophysical methods}

\author[label1]{João B. Fernandes}
\author[label1]{Antônio D. S. Oliveira}
\author[label1]{Mateus C. A. T. Silva}
\author[label2]{Felipe H. Santos-da-Silva}
\author[label1]{Vitor H. M. Rodrigues}
\author[label4]{Kleiton A. Schneider}
\author[label3]{Calebe P. Bianchini}
\author[label1]{João M. de Araujo}
\author[label1]{Tiago Barros}
\author[label2]{Ítalo A. S. Assis}
\author[label1]{Samuel Xavier-de-Souza}
\address[label1]{Universidade Federal do Rio Grande do Norte}
\address[label2]{Universidade Federal Rural do Semi-Árido}
\address[label3]{Instituto Presbiteriano Mackenzie}
\address[label4]{Universidade Federal de Mato Grosso do Sul}

\begin{abstract}
Due to their high computational cost, geophysical applications are typically designed to run in large computing systems. Because of that, such applications must implement several high-performance techniques to use the computational resources better. In this paper, we present Mamute, a software that delivers wave equation-based geophysical methods. Mamute implements two geophysical methods: seismic modeling and full waveform inversion (FWI). It also supports high-performance strategies such as fault tolerance, automatic parallel looping scheduling, and distributed systems workload balancing. We demonstrate Mamute's operation using both seismic modeling and FWI. Mamute is a C++ software readily available under the MIT license.
\end{abstract}


\end{frontmatter}


\begin{table}[!h]
\begin{tabular}{|l|p{6.5cm}|p{6.5cm}|}
\hline
\textbf{Nr.} & \textbf{Code metadata description} & \textbf{Metadata} \\
\hline
C1 & Current code version & 1.0 \\
\hline
C2 & Permanent link to code/repository used for this code version & \url{https://github.com/lapps-ufrn/mamute} \\
\hline
C3  & Permanent link to Reproducible Capsule & \url{https://github.com/lapps-ufrn/mamute/blob/master/doc/mamute/quick-start.md} \\
\hline
C4 & Legal Code License   & MIT License. \\
\hline
C5 & Code versioning system used & git \\
\hline
C6 & Software code languages, tools, and services used & C/C++, Python, OpenMP, MPI. \\
\hline
C7 & Compilation requirements, operating environments \& dependencies & CMake 3.5, GCC 8 \\
\hline
C8 & If available Link to developer documentation/manual & \url{https://github.com/lapps-ufrn/mamute/tree/master/doc} \\
\hline
C9 & Support email for questions & lapps@imd.ufrn.br \\
\hline
\end{tabular}
\caption{Code metadata.}
\label{codeMetadata} 
\end{table}

\section{Motivation and significance} 

Many applications employed by the geophysics community in the search for oil and gas reservoirs are based in the solution of the wave equation. Full waveform inversion (FWI) is one of the main geophysics methods~\cite{Virieux2009}.
It consists of an inverse problem~\cite{Tarantola1984}, wherein seismic data observations are leveraged to glean insights into the Earth’s subsurface. FWI’s inverse problem is solved with numerical optimization methods, which employ the misfit between the observed seismograms and modeled data to obtain its target information. The modeled data in FWI is mostly generated by solving discrete versions of a given wave equation. In this paper, we introduce \texttt{Mamute}, a \texttt{C++} software designed to execute wave equation based methods such as three-dimensional seismic modeling and FWI.

Methods based on the solution of the wave equation
typically require significant time and computational resources.
This is primarily due to the need to apply numerical solutions of partial differential equations (PDEs) and optimization methods on extensive datasets. To enhance performance, \texttt{Mamute} employs parallel programming techniques, leveraging a combination of OpenMP and MPI. Additionally, \texttt{Mamute} is equipped with various tools optimized for modern high-performance computing (HPC) setups. These include workload scheduling techniques for both distributed~\cite{Assis2019} and shared~\cite{Assis2020} memory environments, fault tolerance mechanisms~\cite{Santana2022}, and optimal checkpointing methods~\cite{Symes2007,Griewank2000}, for efficient data modeling and gradient evaluation. These features collectively ensure \texttt{Mamute}'s seamless execution on contemporary supercomputers.

Many of these HPC capabilities have been the focus of numerous research projects. Fernandes et al.~\cite{Fernandes2024} introduced a distributed task scheduling algorithm tailored for \texttt{Mamute}. Da Silva et al.~\cite{Silva2024,patsma2024} explored an auto-tuning approach to schedule OpenMP tasks in \texttt{Mamute} dynamically. Santana et al.~\cite{Santana2021} validated a fault tolerance mechanism within \texttt{Mamute}.

\section{Software description}

\texttt{Mamute} is a C++ program that applies HPC techniques to improve seismic applications. The code is hosted in GitLab and open to contribution with MIT Licence. 

\subsection{Software functionalities}

\texttt{Mamute} is a software suite that implements algorithms based on the numerical solution of the acoustic wave equation in the time domain. \texttt{Mamute} solves the 3D acoustic wave equation through the finite differences method with a second-order approximation in time and an eighth-order approximation in space \cite{LeVeque2007}. Optionally, the wave equation can consider the density influence \cite{DaSilva2022}.
The acoustic wave equation and the variable-density wave equation in a heterogeneous medium are given, respectively, by:
\begin{equation}\label{Eq:Acoustic_Wave}
\frac{\partial^2 p}{\partial t^2}=c^2({\mathbf{x}})\nabla^2p+S({\mathbf{x}},t)\quad\mbox{and}\quad 
\frac{\partial^2 q}{\partial t^2}=c^2({\mathbf{x}})(\nabla^2q-m^2({\mathbf{x}})q)+\frac{S({\mathbf{x}},t)}{\sqrt{\rho({\mathbf{x}})}},
\end{equation}
where ${\mathbf{x}}\in\Omega\subset\mathbb{R}^3$ and $t\in\mathbb{R}$ are, respectively, the Cartesian spatial coordinates and the time, $p({\mathbf{x}},t)$ is the acoustic pressure, $q({\mathbf{x}},t)=p({\mathbf{x}},t)/\sqrt{\rho({\mathbf{x}})}$ is the normalized wavefield, being $\rho({\mathbf{x}})$ the material density, $c({\mathbf{x}})$ is the P-wave velocity of the subsurface, $m^2({\mathbf{x}})=\nabla^2(\sqrt{\rho({\mathbf{x}})})/\sqrt{\rho({\mathbf{x}})}$ represents the mass term. The source term $S({\mathbf{x}},t)$ is often represented as $f(t)\delta({\mathbf{x}}-{\mathbf{x}}_S)$, where $f(t)$ is the time-dependent source signature, ${\mathbf{x}}_S$ denotes the source location, and $\delta({\mathbf{x}}-{\mathbf{x}}_S)$ is the Dirac delta function used to localize the source in space.

Based on that wave propagation operator, \texttt{Mamute} has implemented two seismic processing algorithms, the seismic modeling \cite{Carcione2002} and the full waveform inversion (FWI) \cite{Virieux2009,Tarantola1984}. For the moment, all float-point operations on Mamute are performed with double precision.

\texttt{Mamute}'s seismic modeling features absorbing boundaries implemented via damping \cite{Cerjan1985,Reynolds1978} and convolutional perfectly matched layer (CPML) \cite{RodenGedney2000}. Both techniques are used to allow the simulation of the wave propagation on Earth using a limited computational mesh. 
\texttt{Mamute}'s user can also choose to perform the wave propagation with free surface, \textit{i.e.}, simulating the wave reflections of the surface. Furthermore, time interpolation via the nearest neighbor algorithm or sinc function can be employed to resample seismic data. Using a text configuration file, the user can define the seismic acquisition geometry, including the sources' and receivers' positions, the velocity model's file and its discretization, the source signature file and its peak frequency and amplitude, the absorbing boundary thickness, the number of time steps and the time resolution.

\texttt{Mamute}'s FWI features L-BFGS-b~\cite{Byrd1995,Zhu1997} as its optimization method.
Gradient preconditioning~\citep{trinh2017bessel} is used to smooth the gradient. This software also qualifies for fault tolerance through the Dependability Library for Iterative Applications (DeLIA) \cite{delia2024}, which offers fault detection, data conservation, and failover capabilities. By doing so, the FWI can restart from the state of the execution at a time before the execution is interrupted. The gradient is computed using the adjoint-state method \cite{Plessix2006}. Using a text configuration file, the user can define a limit of memory usage for checkpointing, the L-BFGS-b's boundaries and maximum number of iterations, the maximum number of velocity model updates, and the type of gradient preconditioning (no filter, Bessel filter, or Laplace filter) and its filter coefficients.

The FWI computes the wavefield in forward and reverse time. Forward data needs to be reused when computing the reverse wavefield.
In those circumstances, a high amount of storage is needed to store the forward wavefield. \texttt{Mamute} has three options for managing the forward wavefield. The user can compile \texttt{Mamute} to save the entire wavefield in memory, improving the adjoint performance but consuming a high amount of main memory. For large problems, the RAM may not be enough. An alternative is to save the forward wavefield in the disk. This option will decrease main memory usage considerably, but it will require time to save and read the wavefield from secondary storage. The third choice is to use the optimal checkpointing \cite{Symes2007} technique to save selected timestamps from the forward wavefield and then recompute just the missing data. 

\texttt{Mamute}'s seismic modeling and FWI are implemented with two levels of parallelization. Inter-node parallelization uses the message-passing interface (MPI) \cite{Gropp1999}. The data distribution for inter-node parallelization can be performed either using a static distribution or via cyclic token-based work-stealing (CTWS) \cite{Assis2019}; Intra-node parallelization uses open multi-processing (OpenMP) \cite{Dagum1998}; The OpenMP loop scheduling is auto adjustable using the Parameter Auto-tuning for Shared Memory Algorithms (PATSMA) \cite{patsma2024}.

\subsection{Software architecture}

As previously discussed, \texttt{Mamute} has two main features: Seismic modeling and FWI. Seismic modeling's features are implemented in the \verb|Modeling| class. This class incorporates methods for solving the 3D wave equation. It consists of the most meaningful public methods shown in Table~\ref{tab:model_methods}.

\begin{table}[ht]
\centering
\caption{Description of main Modeling methods.}
\label{tab:model_methods}
\begin{tabularx}{\linewidth}{lX} \hline
\textbf{Method name} & \textbf{Description} \\ \hline
\texttt{ValidConditions} &  Check the Courant–Friedrichs–Lewy condition, a necessary convergence condition for numerically solving PDEs \cite{Carcione2002}. \\
\texttt{Interpolate} & Perform time interpolation of seismic data. \\
\texttt{ReadReceiver} & Read the receivers in a given time step. \\
\texttt{ModelingStep} & Solve the 3D wave equation in a given time step. \\ \hline
\end{tabularx}
\end{table}

The \verb|Modeling| class has attributes of three \texttt{Mamute}'s classes: \verb|Config|, \verb|Grid|, and \verb|VelocityModel|. \verb|Config| is responsible for reading and storing user input parameters. \verb|Grid| stores the grid parameters for the finite difference method (FDM). \verb|VelocityModel| creates the velocity model based on the grid parameters and methods to operate the model. The \verb|ModelingAPP| class uses \verb|Modeling| for performing data modeling and generating synthetic seismograms. Its execution is expressed in Algorithm~\ref{alg:modeling}. \verb|ModelingAPP| reads each seismic source (shot) position and performs the \verb|Forward propagation|. The resulting seismogram from this forward propagation represents the wavefield at specific locations.
The seismic shots can be distributed using MPI communication through the implemented schedulers. Workload scheduling in distributed systems is necessary to balance the workload among nodes to avoid idle time.

\begin{algorithm}[ht]
\caption{Modeling algorithm.}
\label{alg:modeling}
\begin{algorithmic}[1]
\STATE Initialization
\STATE Auto-tuning chunk size of OpenMP loop scheduling \MYCOMMENT{if enabled}
\WHILE{Scheduler returns shot ID}
    \STATE Read shot data
    \STATE Forward propagation
\ENDWHILE
\end{algorithmic}
\end{algorithm}

Regarding FWI, it is composed of three main classes: \verb|FWI|, \verb|Optimizer|, and \verb|Adjoint|. FWI class performs the most general logic of FWI execution expressed in Algorithm~\ref{alg:fwi}. \verb|FWI::run| starts checking the convergence. If the optimization has not converged yet, the optimizer (L-BFGS-b) is executed (Line~\ref{l:fwi:opt}), leading to one of the three following actions: (1) compute a new gradient, (2) update the model using the current gradient using the method \verb|Adjoint::Run|, or (3) set the convergence flag to true.

\begin{algorithm}[ht]
\caption{FWI::run algorithm.}
\label{alg:fwi}
\begin{algorithmic}[1]
\STATE Read initial velocity model and input parameters
\STATE Initiate fault tolerance method \MYCOMMENT{if enabled}
\WHILE{optimizer not converged}
    \STATE Execute optimizer \MYCOMMENT{L-BFGS-b} \label{l:fwi:opt}
    \IF{optimizer requests a new gradient}
        \STATE \texttt{Adjoint::Run} \MYCOMMENT{Algorithm~\ref{alg:adjoint}}
    \ELSIF{optimizer requests to update the model}
        \STATE Model is updated with gradient
    \ELSE 
        \STATE Optimizer has converged
    \ENDIF
    \STATE Save fault tolerance global data \MYCOMMENT{if enabled}
\ENDWHILE
\end{algorithmic}
\end{algorithm}

The method \verb|Adjoint::Run| begins initializing the necessary variables and classes (Algorithm~\ref{alg:adjoint}, Line~\ref{l:adj:init}). A while loop then iterates over a subset of shots for each process (Line~\ref{l:adj:while_start}). If the CTWS scheduler is enabled, it manages the execution of shots for each process; otherwise, the algorithm uses a static scheduler. Following this, the shot data is read (Line~\ref{l:adj:read_shot}). Line~\ref{l:adj:forward_start} marks the initiation of forward propagation, executing \verb|Modeling::ModelingStep| and \verb|Modeling::ReadReceiver|, as outlined in Table~\ref{tab:model_methods}. Additionally, the method \verb|ManageWavefield::SaveWavefield| is invoked. Depending on the forward wavefield management mode, the wavefield will be saved accordingly: to files if using disk storage, to memory buffers if using checkpoints, or entirely in memory if that is the chosen approach.

After the forward propagation, the misfit is calculated, and the seismic data is interpolated. Then, the backward propagation starts (Line~\ref{l:adj:back_start}) where the \verb|Modeling::ModelingStep| is executed. The \texttt{ManageWavefield\\::RetrieveWavefield} recovers the wavefield saved in Line~\ref{l:adj:save_wave}. Then, the image condition is executed using cross-correlation. After the backward propagation finishes, \texttt{ManageWavefield::EndShot} finalize the execution of the shot. After executing all shots, the algorithm uses the MPI to sum up all the process's misfits and gradients.

In the end, the algorithm performs three operations.
\texttt{MultiplyAdjoint} computes the adjoint from the gradient.
The \texttt{GradientePreconditioning} applies a smoothing filter \citep{Trinh2017} to the gradient. Then, \texttt{ZeroesNearSurface} set zeros in the surface to reflect multiplies inside the model.

\begin{algorithm}[!h]
\caption{\texttt{Adjoint::Run} algorithm.}
\label{alg:adjoint}
\begin{algorithmic}[1]
\STATE Initialization \label{l:adj:init}
\STATE Fault tolerance data recovering \MYCOMMENT{if enabled} \label{l:adj:ft_recov}
\STATE Auto-tuning chunk size of OpenMP loop scheduling \MYCOMMENT{if enabled} \label{l:adj:at}
\WHILE{scheduler returns the ID of the next shot} \label{l:adj:while_start}
    \STATE Read shot data \label{l:adj:read_shot}
    \FORALL{Time steps in forward propagation } \label{l:adj:forward_start}
        \STATE Modeling::ModelingStep
        \STATE Modeling::ReadReceiver
        \STATE ManageWavefield::SaveWavefield \label{l:adj:save_wave}
    \ENDFOR \label{l:adj:forward_end}
    \STATE SeismicData::Misfit
    \STATE Modeling::Interpolate
    \FORALL{time steps in backward propagation} \label{l:adj:back_start} 
        \STATE Modeling::ModelingStep
        \STATE ManageWavefield::RetrieveWavefield
        \STATE Image Condition (cross-correlation)
    \ENDFOR
    \STATE ManageWavefield::EndShot
    \STATE Fault tolerance method saves data \MYCOMMENT{if enabled} \label{l:adj:ft_save}
\ENDWHILE \label{l:adj:while_end}
\STATE MPI Allreduce Misfit
\STATE MPI Allreduce Gradient
\STATE MultiplyAdjoint
\STATE GradientPreconditioning
\STATE ZeroesNearSurface
\end{algorithmic}
\end{algorithm}

The \texttt{Adjoint::Run} can also apply HPC techniques such as MPI-based task scheduling, autotuning of parallel loop scheduling, fault tolerance for distributed systems, and memory checkpointing. Fault tolerance and Auto-tuning are features that must be enabled in the compilation. Fault tolerance saves partial data (Line~\ref{l:adj:ft_save}) that can be utilized to recover the execution (Line~\ref{l:adj:ft_recov}) from the last shot in case of failure. Auto-tuning adjusts the chunk size parameter (Line~\ref{l:adj:at}), to be used in the \verb|Modeling::ModelingStep|.

Additionally, an HPC technique is incorporated in the base class \texttt{Manage- Wavefield} that allows various wavefield management modes to be implemented in subclasses. Users have the option to compile \texttt{Mamute} to manage the entire wavefield in memory via the \verb|ManageWavefieldMemory| class, store it on disk with the \verb|ManageWavefieldDisk| class, or create checkpoints using the \texttt{ManageWavefieldChk} class, which works in conjunction with the \verb|Checkpoint| class to run the checkpoint.

\subsubsection{Input parameters file} 
\label{subsec:param_file}

To execute FWI or modeling with \texttt{Mamute}, several parameters must be provided via an input parameter file. Some of these parameters are common to both applications, as outlined in Table~\ref{tab:fwi-model-par}. Parameters \verb|ns|, \verb|dt|, \verb|fpeak|, and \verb|amplitude| should match those used to generate the source signature. Additionally, the \verb|vel| parameter must reference a binary file with double precision. All binary files required for Modeling or FWI must be situated in the \verb|proj_dir| directory.

\begin{table}[!h]
\caption{FWI and Modeling parameters.}
\label{tab:fwi-model-par}
\begin{tabularx}{\linewidth}{lX} \hline
\textbf{Parameter} & \textbf{Description} \\ \hline
\texttt{nx}, \texttt{ny}, \texttt{nz} & Number of points in the velocity model in the x, y, and z dimensions respectively. \\
\texttt{dx}, \texttt{dy}, \texttt{dz} & Distance (in meters) between points in the x, y, and z dimensions respectively. \\
\texttt{border} & Thickness (in points) of the absorbing boundaries. \\
\texttt{ox}, \texttt{oy}, \texttt{oz} & Coordinates (in meters) of the origin in the x, y, and z dimensions respectively. \\
\texttt{ns} & Number of timesteps. \\
\texttt{dt} & Time sampling (in seconds). \\
\texttt{fpeak} & Peak frequency (in hertz). \\
\texttt{amplitude} & Maximum source amplitude. \\
\texttt{n\_src} & Number of sources (shots). \\
\texttt{proj\_dir} & Path to the folder where the input/output binaries are located. \\
\texttt{vel} & Name of the input velocity model data file used for modeling or the initial velocity model for FWI. \\ 
\texttt{density} & Density model file name. \\ 
\hline
\end{tabularx}
\end{table}

Regarding FWI, it has specific parameters listed in Table~\ref{tab:fwi-par}. Users can select the L-BFGS-b boundary type using the parameter \verb|lbfgsb|. Depending on the selected type, two additional parameters must be set: \texttt{lbfgsb\_lower\_bound} if a lower bound is specified, which defines the minimum value that the velocity model can reach during the updates of the L-BFGS-b, and \spverb|lbfgsb_upper_bound| if an upper bound is specified and defines the maximum value. If \spverb|gradient_preconditioning_mode| is set to either Bessel filter or Laplace filter, the filter lengths in the x, y, and z directions can be specified using the parameters \spverb|bessel_filter_lx|, \spverb|bessel_filter_ly|, and \spverb|bessel_filter_lz|, respectively. The default values for these parameters are $2 \times dx$, $2 \times dy$, and $2 \times dz$. \spverb|n_iter| and \spverb|max_viter| are optional parameters. If they are not set, the default value will be $999$ and $100$, respectively.

\begin{table}[!h]
\centering
\caption{FWI exclusive parameters.}
\label{tab:fwi-par}
\begin{tabularx}{\linewidth}{lXl}
\hline
\textbf{Parameter} & \textbf{Description} \\ \hline
\texttt{lbfgsb} & Sets the boundary type in the L-BFGS-b algorithm. It can be unlimited, have only a lower limit, lower and upper limit, or have only an upper limit. \\
\texttt{n\_iter} &  Maximum number of L-BFGS-b iterations. \\
\texttt{max\_viter} & Maximum number of velocity model updates. \\
\texttt{gradient\_preconditioning\_mode} & Method used to smooth the gradient every time it is computed. It can be no preconditioning, Bessel filter, or Laplace filter. \\ \hline
\end{tabularx}
\end{table}

Certain parameters are not directly related to the execution of the \texttt{Mamute} mathematical model but rather to its performance within the environment. Those parameters are detailed in Table~\ref{tab:hpc-par}. \verb|ws_flag| is applicable to both FWI and Modeling applications, whereas the others are exclusive to FWI. It is important to note that \verb|check_mem| and \verb|chk_verb| are optional, with default values of $0.8$ and $0$, respectively. When \texttt{Mamute} is compiled with \verb|VERBOSE=ON|, the parameter \verb|chk_verb| comes into play - setting it to $1$ prompts \texttt{Mamute} to write checkpointing buffers to \verb|chkbuf.bin|, while setting it to $0$ prevents writing. Furthermore, \verb|ft_config| and \verb|fwi_config| should be initialized to enable fault tolerance if \texttt{Mamute} has been compiled with \verb|FT=ON|. See \cite{delia2024} for more details.

\begin{table}[!h]
\centering
\caption{Mamute HPC parameters.}
\label{tab:hpc-par}
\begin{tabularx}{\linewidth}{lXl}
\hline
\textbf{Parameter} & \textbf{Description} \\ \hline
\texttt{check\_mem} & Value between (0,1] that represents the percentage of memory usage by checkpointing buffers. \\
\texttt{chk\_verb} & Checkpointing verbosity level. \\
\texttt{ws\_flag} & Flag used to active CTWS. \\
\texttt{ft\_config} & Path to file with DeLIA parameters for fault tolerance. \\
\texttt{fwi\_config} & Path to file with FWI parameters. \\ \hline
\end{tabularx}
\end{table}

\subsubsection{Data format}
\label{subsec:data}
To execute \texttt{Mamute}, some input files are required. All of them are binary files, except for the configuration file, and must be stored in the \verb|proj_dir| directory. Table~\ref{tab:files} list those files. All of them have a predetermined filename, except for the velocity model that needs to be defined using the \texttt{vel} parameter in the configuration file.

\begin{table}[!ht]
\centering
\caption{List of files inside \texttt{proj\_dir} and the relation (input or output) to Modeling and FWI. \centering}
\label{tab:files}
\begin{tabular}{lcc}
\hline
\textbf{File} & \textbf{Modeling} & \textbf{FWI} \\ \hline
\texttt{rcv\_coord\_i.bin} & input & input \\
\texttt{source.bin} & input & input \\
\texttt{src\_coord.bin} & input & input \\
\texttt{vel} & input & input \\
\texttt{dobs\_i.bin} & output & input \\
\texttt{v-final.bin} &  & output \\
\texttt{v-iter-k.bin} &  & output \\ \hline
\end{tabular}
\end{table}

Each shot in \texttt{Mamute} has two specific files associated with it: \texttt{rcv\_coord\_i.bin} for receiver coordinates and \verb|dobs_i.bin| for observed data (seismogram), where $i$ represents the shot number. The receiver file should contain the 3D coordinates of the receivers in the order of $(x,y,z)$. These coordinates should correspond with the order of the traces in the observed data. This means that the initial $ns \times \text{double}$ bytes in \verb|dobs_i.bin| will correspond to the $ns$ samples of the first trace of shot $i$.

The source signature is stored in the \verb|source.bin| file, which contains $ns$ samples, i.e., its size must be $ns$ times the size of a \texttt{double} in bytes. The source coordinates must be stored in the \verb|src_coord.bin| file, which contains one 3D coordinate per shot sorted by the shot IDs. Each 3D coordinate is represented in meters in the order $(x,y,z)$.

\texttt{Mamute} supports initialization with a velocity model from a binary file. The name of the velocity model should be specified in parameter \verb|vel| within the configuration file. The 3D velocity model must be stored in the file using the sequence of dimensions: $z \times y \times x$. Additionally, \texttt{Mamute} offers scripts for creating synthetic velocity models, such as a constant model, a centered sphere, and a centered sphere with Gaussian perturbation.

Also, Mamute supports using a binary file for the density model. The file name must be specified in the \verb|density| parameter within the configuration file. Alternatively, the user can use a script provided by Mamute to generate a density model from the velocity model using Gardner's equation \cite{Gardner1974}.

The modeling output consists of a set of seismograms that follow the same format as the observed data. In the case of Full Waveform Inversion (FWI), the primary output is a model named \verb|v-final.bin|. Additionally, \texttt{Mamute} provides the output model for each FWI's iteration as \verb|v-iter-k.bin|, where $k$ represents the FWI iteration number. Each model is saved in a file following the sequence of dimensions: $z \times y \times x$.

\section{Illustrative example}

\begin{figure}[h]
    \centering
    \begin{subfigure}[b]{0.48\textwidth}
        \centering
        \includegraphics[width=\textwidth]{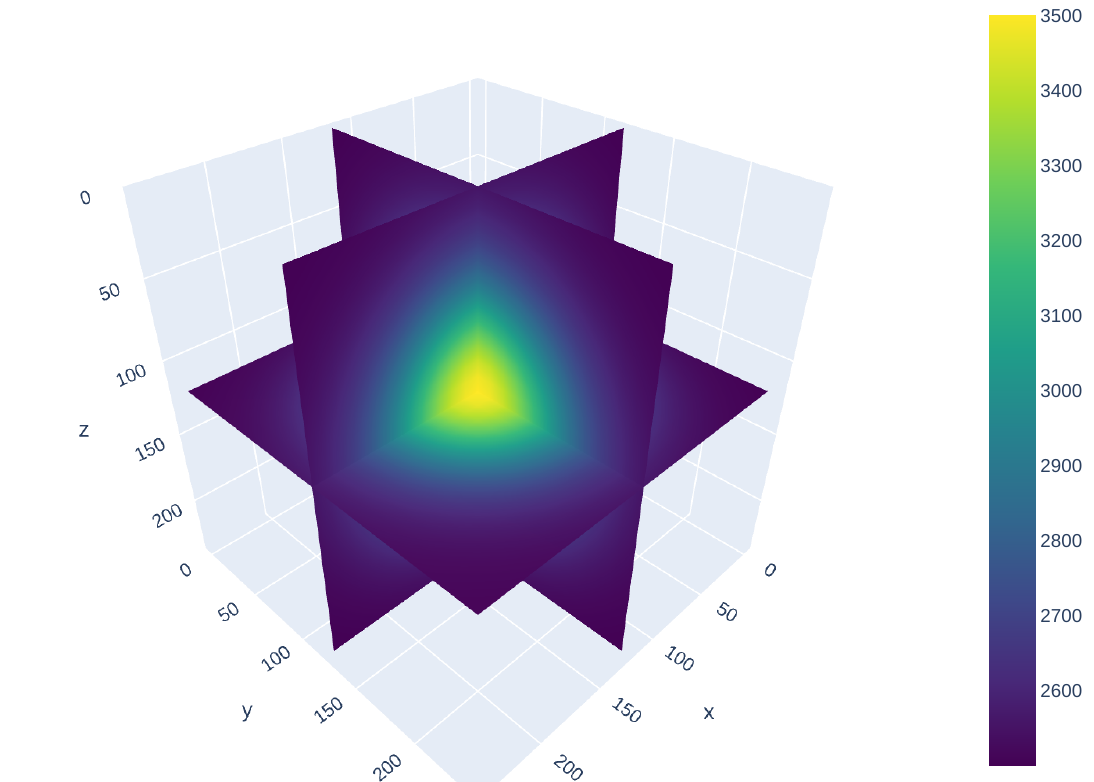}
        \caption{}
        \label{fig:vel-model}
    \end{subfigure}
    \hfill
    \begin{subfigure}[b]{0.48\textwidth}
        \centering
        \includegraphics[width=\textwidth]{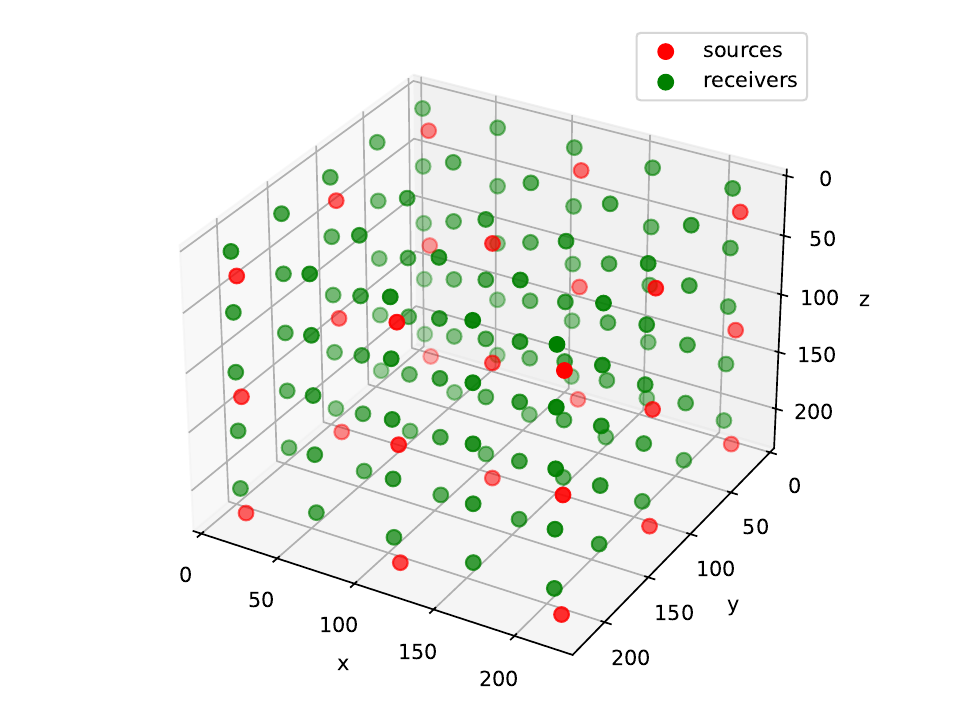}
        \caption{}
        \label{fig:src-rcv}
    \end{subfigure}
    \caption{(a) True velocity model and (b) acquisition geometry, source, and receiver coordinates, used in the Modeling and FWI examples.}
\end{figure}

This section presents an FWI example of \texttt{Mamute} using synthetic data. Figure~\ref{fig:vel-model} shows the true model for this example. For each coordinate $(i,j,k)$, the true velocity model is calculated using the following equation 
\begin{equation}
    v(i, j, k) = v_{\mathrm{max}} + 10^3 \cdot \exp\left(-0.5 \cdot \frac{(i - \frac{nx}{2})^2 + (j - \frac{ny}{2})^2 + (k - \frac{nz}{2})^2}{\sigma^2} \right),
\end{equation}
where $v_{\mathrm{max}} = 2500\text{m/s}$ is the maximum velocity, and $nx = ny = nz = 25$ are the number of points on dimensions $x$, $y$, and $z$, respectively. The space between points for each dimension is $10$ meters. 

\begin{figure}
    \centering
    \includegraphics[width=0.7\linewidth]{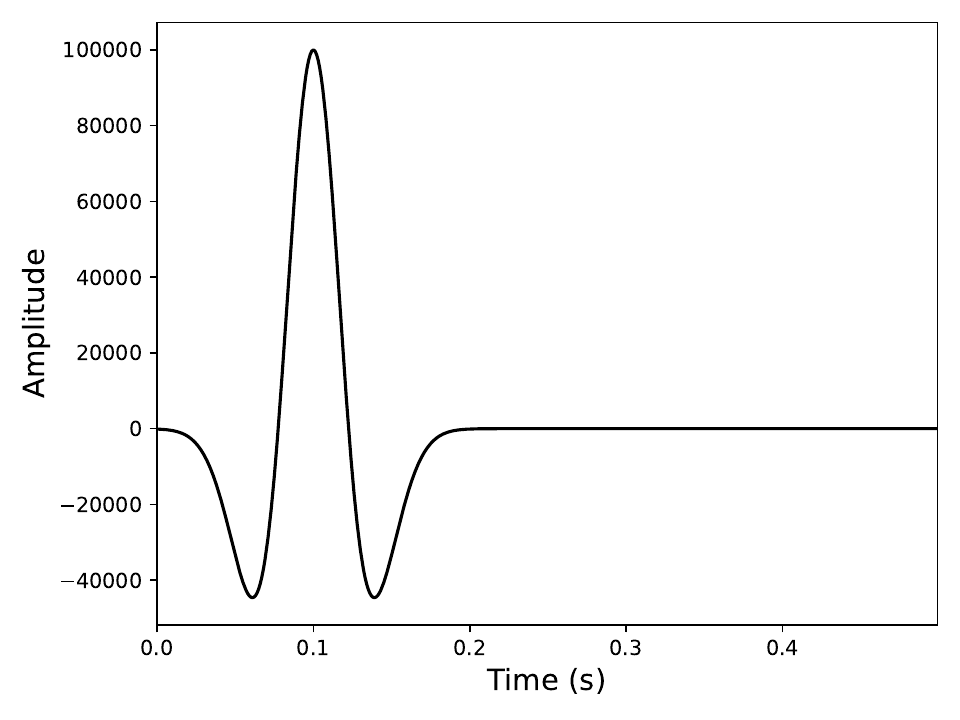}
    \caption{Ricker source wavelet.}
    \label{fig:source}
\end{figure}

The observed data in this example are generated using \texttt{Mamute}'s modeling application, employing the true velocity model illustrated in Figure~\ref{fig:vel-model}. We set the sources and receiver's positions as shown in Figure~\ref{fig:src-rcv}. The receivers are positioned at intervals of $50 \text{m}$, ranging from $10 \text{m}$ to $210 \text{m}$. In contrast, the sources are spaced at $100 \text{m}$ intervals, from $20 \text{m}$ to $220 \text{m}$.
As shown in Figure~\ref{fig:source}, a Ricker function is used as the source wavelet. The configuration file for the modeling application in this example is shown below.
\begin{verbatim}
    nx = 25
    ny = 25
    nz = 25
    ns = 500
    border = 25
    dx = 10.0
    dy = 10.0
    dz = 10.0
    vel = velocity_model.bin
    fpeak = 10
    dt = 0.001
    amplitude = 100000
    n_src = 27
    proj_dir = ./projects/example
    stencil = 4
    ox = 0.0
    oy = 0.0
    oz = 0.0
\end{verbatim}

\begin{figure}[!ht]
    \centering
    \includegraphics[width=\linewidth]{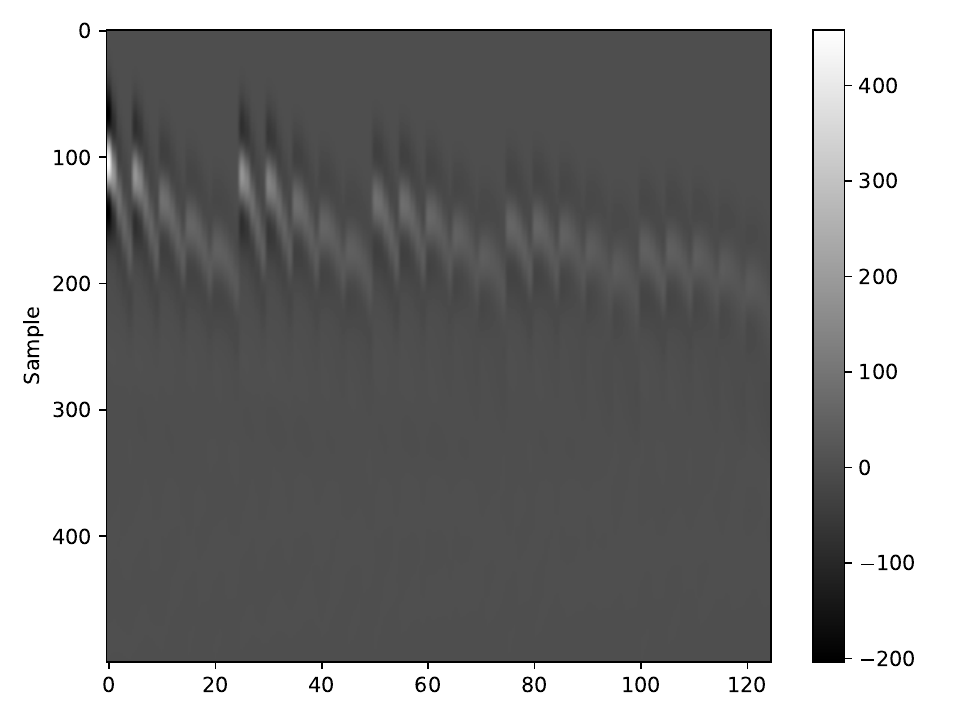}
    \caption{Seismogram for the shot $0$.}
    \label{fig:seism_0}
\end{figure}

To run \texttt{Mamute}'s modeling application, reading the input parameters from the file \verb|config_modeling.txt|, the user can execute the following command: 
\begin{verbatim}
    mpirun -n <np> modeling config_modeling.txt
\end{verbatim}
where \verb|<np>| is the number of processes. After the code execution is complete, several files containing the observed data are created. The data format was previously explained in Section~\ref{subsec:data}. Figure~\ref{fig:seism_0} displays an example of observed data for the first shot of the simulation.

After the observed data, the user can proceed with the test and run \texttt{Mamute}'s FWI application. Here, the observed data are used as input for the execution. The configuration parameters are also read from the file. We extend the \verb|config_modeling.txt| by adding the following lines:
\begin{verbatim}
    [...] // Repeat Modeling parameters
    lbfgsb = 2
    lbfgsb_lower_bound = 2000.0
    lbfgsb_upper_bound = 3500.0
    n_iter = 10
    max_viter = 5
    gradient_preconditioning_mode = 2
    zeroes_nplanes_gradient = 0
\end{verbatim}
Additionally, the option \verb|vel| is modified to \verb|velocity_initial.bin|, which refers to the initial velocity model file, as depicted in Figure~\ref{fig:vinit}. This creates the \verb|config_fwi.txt| file.

Then, employing the defined parameter file, the user can run the \texttt{Mamute}'s FWI application using the following command:
\begin{verbatim}
    mpirun -n <np> fwi config_fwi.txt
\end{verbatim}

\begin{figure}[h]
    \centering
    \begin{subfigure}[b]{0.48\textwidth}
        \centering
        \includegraphics[width=\textwidth]{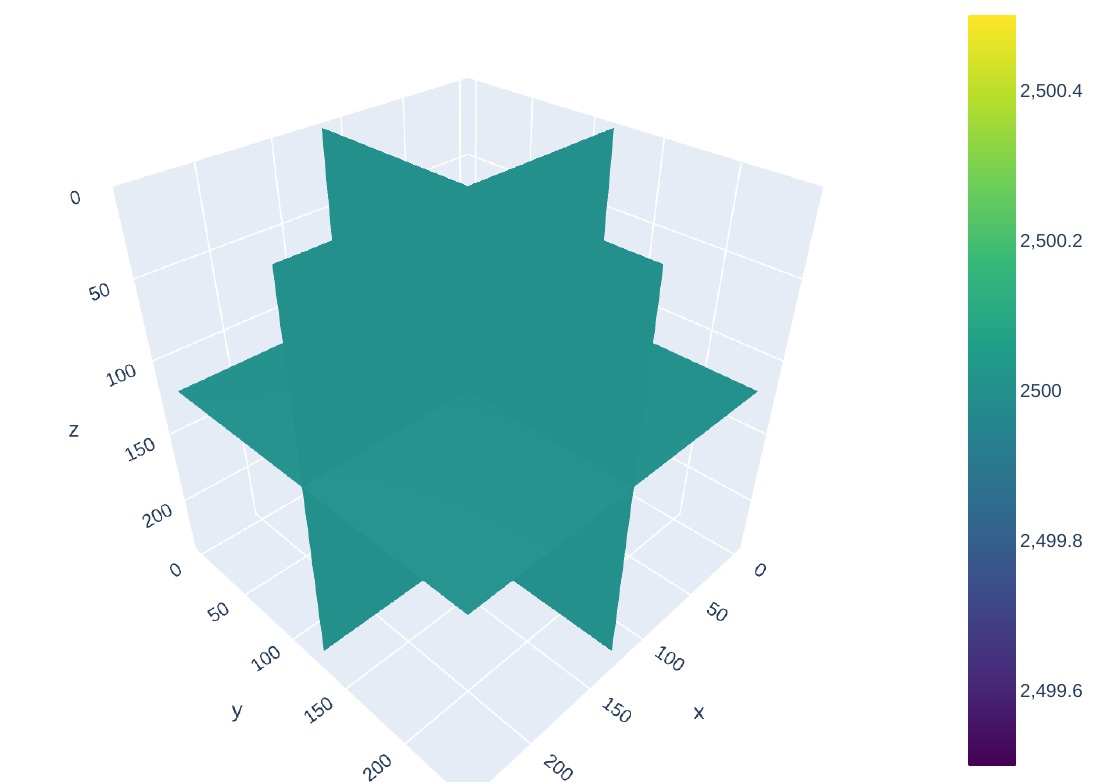}
        \caption{}
        \label{fig:vinit}
    \end{subfigure}
    \hfill
    \begin{subfigure}[b]{0.48\textwidth}
        \centering
        \includegraphics[width=\textwidth]{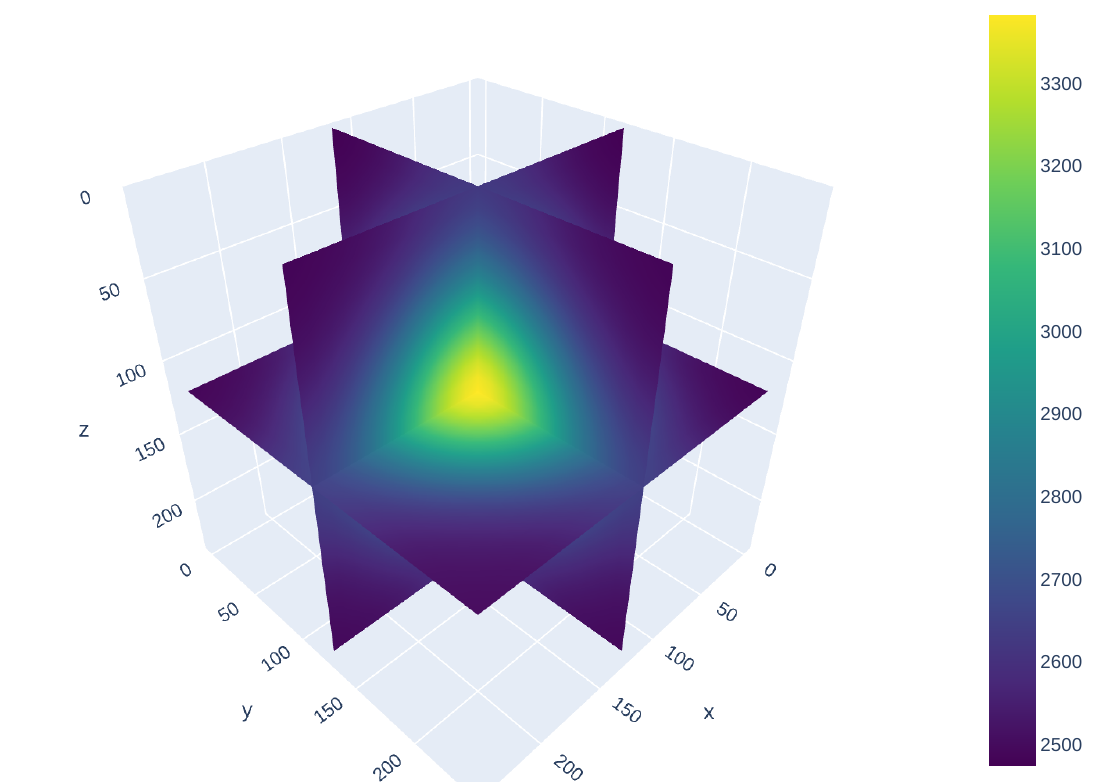}
        \caption{}
        \label{fig:vfinal}
    \end{subfigure}
    \caption{Initial (a) and final (b) velocity models from \texttt{Mamute}'s FWI execution.}
\end{figure}

After the execution is complete, the output of the FWI will contain the final velocity model, shown in Figure~\ref{fig:vfinal}. A detailed version of this example is available in the tutorials in \texttt{Mamute}'s repository.

\section{Impact}

\texttt{Mamute} is a \texttt{C++} tool that implements 3D seismic modeling and FWI algorithms based on 3D wave equations. Seismic modeling techniques are primarily used to generate synthetic seismic data, which is useful for investigating potential areas of interest in oil and gas exploration. FWI techniques are employed to estimate seismic velocity models. These models are used to image seismic data through migration techniques. These methods are primarily applied in oil and gas exploration through seismic data processing and reservoir monitoring via time-lapse seismic methods.

Since the implementation of wave equation techniques is highly computing-intensive, \texttt{Mamute} offers several high-performance computing (HPC) techniques to improve performance during large-scale execution and enable \texttt{Mamute} to run on supercomputers.

\texttt{Mamute} has been publicly released since March 2024. The authors have already utilized Mamute in several published studies \cite{Assis2019, Assis2020, Fernandes2024, Silva2024, delia2024}.

\section*{Acknowledgements}
\label{}
\textit{The authors gratefully acknowledge support from Shell Brazil through the project ``\textit{Novas metodologias computacionalmente escal\'{a}veis para s\'{i}smica 4D orientado ao alvo em reservat\'{o}rios do pr\'{e}-sal}'' and Equinor through the project ``Non-Stationary Deconvolution and Imaging
for High Resolution Seismic'' at the Universidade Federal do Rio Grande do Norte (UFRN) and the strategic importance of the support given by ANP through the R\&D levy regulation.
The authors also acknowledge the National Laboratory for Scientific Computing (LNCC/MCTI, Brazil) and the High-Performance Computing Center at UFRN (NPAD/UFRN) for providing HPC resources of the SDumont and NPAD supercomputers. Felipe Silva has been an undergraduate researcher through the PIVIC and PIVIC-Af programs at UFERSA.}



\bibliographystyle{elsarticle-num} 
\bibliography{references}

\begin{thebibliography}{10}
\expandafter\ifx\csname url\endcsname\relax
  \def\url#1{\texttt{#1}}\fi
\expandafter\ifx\csname urlprefix\endcsname\relax\def\urlprefix{URL }\fi
\expandafter\ifx\csname href\endcsname\relax
  \def\href#1#2{#2} \def\path#1{#1}\fi

\bibitem{Virieux2009}
J.~Virieux, S.~Operto, An overview of full-waveform inversion in exploration
  geophysics, GEOPHYSICS 74 (2009) WCC1--WCC26.
\newblock \href {https://doi.org/10.1190/1.3238367}
  {\path{doi:10.1190/1.3238367}}.

\bibitem{Tarantola1984}
A.~Tarantola, Inversion of seismic reflection data in the acoustic
  approximation, GEOPHYSICS 49 (1984) 1259--1266.
\newblock \href {https://doi.org/10.1190/1.1441754}
  {\path{doi:10.1190/1.1441754}}.

\bibitem{Assis2019}
I.~A.~S. Assis, A.~D.~S. Oliveira, T.~Barros, I.~M. Sardina, C.~P. Bianchini,
  S.~X. De-Souza, Distributed-memory load balancing with cyclic token-based
  work-stealing applied to reverse time migration, IEEE Access 7 (2019)
  128419--128430.
\newblock \href {https://doi.org/10.1109/ACCESS.2019.2939100}
  {\path{doi:10.1109/ACCESS.2019.2939100}}.

\bibitem{Assis2020}
I.~A.~S. Assis, J.~B. Fernandes, T.~Barros, S.~Xavier-De-Souza, Auto-tuning of
  dynamic scheduling applied to 3d reverse time migration on multicore systems,
  IEEE Access 8 (2020) 145115--145127.
\newblock \href {https://doi.org/10.1109/ACCESS.2020.3015045}
  {\path{doi:10.1109/ACCESS.2020.3015045}}.

\bibitem{Santana2022}
C.~Santana, I.~Sardina, S.~Xavier-De-Souza, Fault tolerance library for
  geophysical numerical methods, 3rd EAGE Workshop on HPC in Americas 2022
  (2022) 1--5.
\newblock \href {https://doi.org/10.3997/2214-4609.2022.80004/CITE/REFWORKS}
  {\path{doi:10.3997/2214-4609.2022.80004/CITE/REFWORKS}}.

\bibitem{Symes2007}
W.~W. Symes, Reverse time migration with optimal checkpointing, GEOPHYSICS 72
  (2007) SM213--SM221.
\newblock \href {https://doi.org/10.1190/1.2742686}
  {\path{doi:10.1190/1.2742686}}.

\bibitem{Griewank2000}
A.~Griewank, A.~Walther, Algorithm 799: revolve, ACM Transactions on
  Mathematical Software 26 (2000) 19--45, análise da quanitidade ótima de
  checkpointings.
\newblock \href {https://doi.org/10.1145/347837.347846}
  {\path{doi:10.1145/347837.347846}}.

\bibitem{Fernandes2024}
J.~B. Fernandes, Ítalo A. S.~de Assis, I.~M.~S. Martins, T.~Barros, S.~X.
  de~Souza, \href{https://arxiv.org/abs/2401.04494v2}{Adaptive asynchronous
  work-stealing for distributed load-balancing in heterogeneous systems}, arXiv
  (1 2024).
\newline\urlprefix\url{https://arxiv.org/abs/2401.04494v2}

\bibitem{Silva2024}
F.~H.~S. da~Silva, J.~B. Fernandes, I.~M. Sardina, T.~Barros, S.~X. de~Souza,
  I.~A.~S. Assis, \href{https://arxiv.org/abs/2402.16728v1}{Auto tuning for
  openmp dynamic scheduling applied to fwi}, arXiv (2 2024).
\newline\urlprefix\url{https://arxiv.org/abs/2402.16728v1}

\bibitem{patsma2024}
J.~B. Fernandes, F.~H.~S. da~Silva, T.~Barros, I.~A. Assis, S.~X. de~Souza,
  Patsma: Parameter auto-tuning for shared memory algorithms, SoftwareX 27
  (2024) 101789.
\newblock \href {https://doi.org/https://doi.org/10.1016/j.softx.2024.101789}
  {\path{doi:https://doi.org/10.1016/j.softx.2024.101789}}.

\bibitem{Santana2021}
C.~Santana, I.~Assis, T.~Barros, I.~Sardina, S.~X. de~Souza, Fault tolerance
  applied to 3d full waveform inversion, in: Proceedings of the Digital
  Subsurface Conference in Latin America, Vol. 2021, European Association of
  Geoscientists \& Engineers, 2021, pp. 1--5.
\newblock \href {https://doi.org/10.3997/2214-4609.202181013}
  {\path{doi:10.3997/2214-4609.202181013}}.

\bibitem{LeVeque2007}
R.~J. LeVeque, Finite Difference Methods for Ordinary and Partial Differential
  Equations, Society for Industrial and Applied Mathematics, 2007.
\newblock \href {https://doi.org/10.1137/1.9780898717839}
  {\path{doi:10.1137/1.9780898717839}}.

\bibitem{DaSilva2022}
S.~Da~Silva, A.~Karsou, R.~Moreira, J.~Lopez, M.~Cetale, Klein-gordon equation
  and variable density effects on acoustic wave propagation in brazilian
  pre-salt fields, EarthDoc 2022~(1) (2022) 1--5.
\newblock \href {https://doi.org/https://doi.org/10.3997/2214-4609.202210385}
  {\path{doi:https://doi.org/10.3997/2214-4609.202210385}}.

\bibitem{Carcione2002}
J.~M. Carcione, G.~C. Herman, A.~P.~E. ten Kroode, Seismic modeling, GEOPHYSICS
  67 (2002) 1304--1325.
\newblock \href {https://doi.org/10.1190/1.1500393}
  {\path{doi:10.1190/1.1500393}}.

\bibitem{Cerjan1985}
C.~Cerjan, D.~Kosloff, R.~Kosloff, M.~Reshef, A nonreflecting boundary
  condition for discrete acoustic and elastic wave equations (1985).
\newblock \href {https://doi.org/10.1190/1.1441945}
  {\path{doi:10.1190/1.1441945}}.

\bibitem{Reynolds1978}
A.~C. Reynolds, Boundary conditions for the numerical solution of wave
  propagation problems, Geophysics 43 (1978) 1099--1110.
\newblock \href {https://doi.org/10.1190/1.1440881}
  {\path{doi:10.1190/1.1440881}}.

\bibitem{RodenGedney2000}
J.~A. Roden, S.~D. Gedney, Convolution pml (cpml): An efficient fdtd
  implementation of the cfs–pml for arbitrary media, Microwave and Optical
  Technology Letters 27~(5) (2000) 334--339.
\newblock \href
  {https://doi.org/https://doi.org/10.1002/1098-2760(20001205)27:5<334::AID-MOP14>3.0.CO;2-A}
  {\path{doi:https://doi.org/10.1002/1098-2760(20001205)27:5<334::AID-MOP14>3.0.CO;2-A}}.

\bibitem{Byrd1995}
R.~H. Byrd, P.~Lu, J.~Nocedal, C.~Zhu, A limited memory algorithm for bound
  constrained optimization, SIAM Journal on Scientific Computing 16 (1995)
  1190--1208.
\newblock \href {https://doi.org/10.1137/0916069} {\path{doi:10.1137/0916069}}.

\bibitem{Zhu1997}
C.~Zhu, R.~H. Byrd, P.~Lu, J.~Nocedal, Algorithm 778: L-bfgs-b: Fortran
  subroutines for large-scale bound-constrained optimization, ACM Trans. Math.
  Softw. 23 (1997) 550--560.
\newblock \href {https://doi.org/10.1145/279232.279236}
  {\path{doi:10.1145/279232.279236}}.

\bibitem{trinh2017bessel}
P.-T. Trinh, R.~Brossier, L.~M{\'e}tivier, J.~Virieux, P.~Wellington, Bessel
  smoothing filter for spectral-element mesh, Geophysical Journal International
  209~(3) (2017) 1489--1512.

\bibitem{delia2024}
C.~Santana, R.~C. Araújo, I.~M. Sardina, Ítalo A.S.~{de Assis}, T.~Barros,
  C.~P. Bianchini, A.~D. de~S.~Oliveira, J.~M. {de Araújo}, H.~Chauris,
  C.~Tadonki, S.~X. de~Souza, Delia: A dependability library for iterative
  applications applied to parallel geophysical problems, Computers \&
  Geosciences (2024) 105662\href
  {https://doi.org/https://doi.org/10.1016/j.cageo.2024.105662}
  {\path{doi:https://doi.org/10.1016/j.cageo.2024.105662}}.

\bibitem{Plessix2006}
R.~E. Plessix, A review of the adjoint-state method for computing the gradient
  of a functional with geophysical applications, Geophysical Journal
  International 167 (2006) 495--503.
\newblock \href {https://doi.org/10.1111/j.1365-246X.2006.02978.x}
  {\path{doi:10.1111/j.1365-246X.2006.02978.x}}.

\bibitem{Gropp1999}
W.~Gropp, E.~Lusk, A.~Skjellum, Using MPI, 2nd Edition: Portable Parallel
  Programming with the Message Passing Interface, MIT Press, 1999.

\bibitem{Dagum1998}
L.~Dagum, R.~Menon, Openmp: an industry standard api for shared-memory
  programming, IEEE Computational Science and Engineering 5 (1998) 46--55.
\newblock \href {https://doi.org/10.1109/99.660313}
  {\path{doi:10.1109/99.660313}}.

\bibitem{Trinh2017}
P.~Trinh, R.~Brossier, L.~Métivier, J.~Virieux, P.~Wellington, Bessel
  smoothing filter for spectral-element mesh, Geophysical Journal International
  209~(3) (2017) 1489--1512.
\newblock \href {https://doi.org/10.1093/gji/ggx103}
  {\path{doi:10.1093/gji/ggx103}}.

\bibitem{Gardner1974}
G.~H.~F. Gardner, L.~W. Gardner, A.~R. Gregory, Formation velocity and
  density—the diagnostic basics for stratigraphic traps, GEOPHYSICS 39~(6)
  (1974) 770--780.
\newblock \href {https://doi.org/10.1190/1.1440465}
  {\path{doi:10.1190/1.1440465}}.

\end{thebibliography}






\end{document}